\documentclass[a4paper,12pt]{article}
\usepackage[cp1251]{inputenc}
\usepackage[english]{babel}
\usepackage[dvips]{graphicx}
\graphicspath{{.}}
\usepackage{amsfonts}
\usepackage{amsmath}

\begin{document}
\mathsurround=2pt \sloppy
\title{\bf Analog of Anderson theorem for the polar phase of liquid $^3$He in nematic aerogel}
\author{ I. A. Fomin
\vspace{.5cm}\\
{\it P. L. Kapitza Institute for Physical Problems}\\ {\it Russian
Academy of Science},\\{\it Kosygina 2,
 119334 Moscow, Russia}}
\maketitle
\begin{abstract}
It is shown that if impurities in superfluid $^3$He have form of infinitely long non-magnetic strands, which are straight, parallel to each other and  reflect quasi-particles specularly, the temperature of transition of liquid  $^3$He  from the normal into the polar phase coincides with that for the bulk liquid without impurities. Magnetic scattering lowers transition temperature for the polar phase in analogy with the Anderson theorem for conventional superconductors. These results are discussed in connection with the recent experimental findings.
\end{abstract}

\section{Introduction}
The statement of the theory of super-conducting alloys \cite{AG2}, known as Anderson theorem \cite{And}, applies to conventional  (s-wave) superconductors.
According to this statement elastic scattering of Fermi quasi-particles by non-magnetic impurities does not lower the temperature $T_c$ of the transition  in the super-conducting or superfluid state.  On the contrary, if the pairing is unconventional even non-magnetic impurities lower T$_c$ \cite{lark,AG3}. A relative value of the lowering $\delta T_c/T_c$ is of the order of $\xi_0/\lambda$, where $\xi_0$ is the coherence length of a superconductor and $\lambda$ - the corresponding  mean free path. In superfluid $^3$He the $p$-wave Cooper pairing is realized. The order parameter is the 2$\times$2 spin matrix $\Delta_{\alpha \beta}=\Delta \vec{d}(\hat{k})(i\vec{\sigma}\sigma_y)_{\alpha \beta}$, where $\Delta$ is an overall amplitude, $\vec{d}(\hat{k})$ - vector function of the direction $\hat{k}$ in the momentum space and $\sigma_\zeta$ are Pauli matrices. For the $p$-wave pairing $d_{\mu}(\hat{k})=A_{\mu j}\hat{k})_j$, where $A_{\mu j}$ is a complex 3$\times$3 matrix. Different superfluid phases have different  $A_{\mu j}$. For the polar phase $A_{\mu j}=\exp(i\varphi)d_{\mu} m_j$, where $d_{\mu}$ is a real unit vector in spin space and $m_j$ - a real unit orbital vector.

 Part of impurities in $^3$He is played by strands of aerogel. In the most of the experiments with the superfluid $^3$He high-porosity aerogels are used, which occupy $\approx$2$\%$ of the volume of a sample, so that the mean free path of single-particle excitations is much greater than the interatomic distance in liquid $^3$He and this condition of applicability of Abrikosov and Gorkov (AG) \cite{AG1,AG2} theory of super-conducting alloys is met.
 On the other hand the scattering of excitations by the strands of aerogel is different from the scattering by typical atomic impurities in metals. Diameters $d$ of strands for different aerogels are in a range of 3$\div$10 nm, which is much greater than the Fermi wavelength. If a  strand is represented as a thread consisting of "elementary scatterers" these scatterers can not be considered as point-like. Moreover, they are anisotropic and  have to be characterized by their orientations. Aoyama and Ikeda \cite{AI} have discussed possible effect of oriented anisotropic  impurities on the phase diagram of a p-wave superfluid. They have modeled average square of the Born scattering amplitude  as $\overline{|u_{\mathbf{k}}|^2}=A[1+\delta_u(\hat{\mathbf{k}}\cdot\hat{z})^2]$. Parameter $\delta_u$, which is assumed to be small in their argument, measures sign and magnitude of the global anisotropy. Small anisotropy can be achieved by a deformation of originally isotropic aerogel. On the basis of their calculations  Aoyama and Ikeda arrived at the stimulating conclusion that at a negative $\delta_u$ (in their paper $\delta_u=-0.07$), which corresponds to stretched aerogel a region of stability of the polar phase appears on the phase diagram of liquid  $^3$He. The order parameter of this phase has a form  $A_{\mu j}=\exp(i\varphi)d_{\mu} m_j$ where $d_{\mu}$ is a real unit spin vector and $m_j$ is the unit orbital vector, oriented along the axis of global anisotropy $z$. The global anisotropy lowers symmetry of the system from spherical to axial and  splits the $p$-wave transition temperature in two,   corresponding to different projections of angular momentum $l_z=0$ and $l_z=\pm 1$. The polar phase is that of $l_z=0$. The polar phase was first observed in the experiments \cite{Dm-pp}. For its observation Dmitriev et al. used a special, nematically ordered aerogel - "nafen" \cite{nf}. Its strands are very long and practically parallel to each other. In the experiments \cite{Dm-pp} the strands of nafen were covered by $\approx$2.5 atomic layers of  $^4$He to prevent formation of paramagnetic solid  $^3$He on the strands. When the strands are covered by a film of $^4$He the scattering of quasi-particles can be considered as specular at pressures below 20 \emph{bar}, while without the film it is diffuse at all pressures \cite{kim}.
 Except for that an exchange of spins between quasi-particles and the adsorbed atoms of  $^3$He becomes possible.
 Recent experiments  of Dmitriev et al.\cite{Dm-ms} have shown that the coverage of strands by a sufficiently thick (approximately 2.5 layers) film of  $^4$He  is a crucial condition of  observation of the polar phase. Without the film the phase diagram is essentially different.

The present paper is stimulated by the results of the experiments \cite{Dm-ms}. Its goal is to clarify a question why the specular reflection of the excitations by the strands of nafen is so important for stabilization of the
polar phase of $^3$He. For that purpose an idealized model of  $^3$He in nafen is considered. It is represented as a set of infinitely long thin cylinders oriented parallel to each other and randomly distributed with the average density $n_2$ in the plane $x,y$, perpendicular to the axis of anisotropy  $z$. It is assumed that the surface of the cylinders is sufficiently smooth to secure the specular reflection of quasi-particles.  For this idealized model the temperature of transition from the normal to the polar phase is found. The width of the temperature interval where the polar phase is stabilized is found too.  Effect of the magnetic (exchange) scattering of excitations by the adsorbed  $^3$He on the phase diagram is estimated.

\section{Theorem}
Following the AG theory of super-conducting alloys we describe interaction of quasi-particles with the strands of aerogel by the static potential
$U(\mathbf{r})=\sum_a u(\rho-\rho_a)$. Here  $\rho=(x,y)$ is a two-dimensional  vector. Summation is going over the coordinates $x_a, y_a$ of projections of  strands on the plane  $x, y$.   Fourier transform of the potential is
$$
U(\mathbf{k})=2\pi\delta(k_z)u(\kappa)\sum_a e^{-i\kappa\rho_a},                              \eqno(1)
$$

where $\kappa=(k_x,k_y)$ is a two-dimensional wave vector, $u(\kappa)=\int u(\rho)\exp(i\kappa\rho)d^2\rho$. Presence of $\delta(k_z)$ in the r.h.s. of Eq. () means that at an elementary process of scattering of quasi-particles
not only energy but the longitudinal projection of momenta is conserved. This is a formal expression of the assumption of the specular reflection by the strands. Possible anisotropy of pairing interaction is not considered here and  The interaction of quasi-particles leading to the \emph{p}-wave Cooper pairing is taken in the form $V(\mathbf{k},\mathbf{k^{\prime}})=3g(\mathbf{k}\cdot\mathbf{k^{\prime}})$.
 In the bulk  $^3$He transition temperature T$_c^{(0)}$ for all components of $l$ is the same,  In particular for $l_z=0$ it is found from the equation
$$
1=-3\lambda T^0_c\sum_n\int k_z^2G^{(0)}(\omega_n\mathbf{k})G^{(0)}(-\omega_n\mathbf{-k})\frac{d^3k}{(2\pi)^3},        \eqno(2)
$$
where $G^{(0)}(\omega_n\mathbf{k})$ is the one-particle Green function in momentum representation:
$G^{(0)}(\omega_n,\mathbf{k})=(i\omega_n -\xi+i\delta sgn \omega_n)^{(-1)}$.
For account of the effect of the random static field (1) the standard procedure of averaging over realizations of the field was used. The field $U(\mathbf{r})$ is considered as a perturbation. On the average it has axial symmetry so that the states with different projections of orbital momenta on the symmetry axis $l_z=0$ and $l_z=\pm 1$ form correct zero-order basis and can be considered separately. It is assumed also, that concentration of impurities is small and it does not change significantly the constant $g$ of interaction of excitations.

We start with the $l_z=0$ state. Further terms of the perturbation series contain two types of essential contributions.
First, the denominators of the Green functions acquire self energy terms $\Sigma(\varepsilon,\kappa)$. The first order contribution $\Sigma^{(1)}=n_2\int u(\rho)d^2\rho$, where $n_2$ is the two-dimensional density of strands, is absorbed in the chemical potential. The second order term has to be found self-consistently from the equation:
$$
\Sigma^{(2)}(\varepsilon,\kappa)=n_2\int\frac{d^2\kappa'}{(2\pi)^2}\frac{|u(\kappa-\kappa')|^2}{\varepsilon-\xi_{\kappa}-\Sigma^{(2)}(\varepsilon,\kappa')+i\delta  sgn \varepsilon}.        \eqno(3)
$$
This equation is a 2D analog of the corresponding 3D equation, $\xi_{\kappa}=\varepsilon(\mathbf{k})-\mu$, but $k_z$ is fixed. Then at the transition in the integral Eq. (3) to integration over $\xi_{\kappa}$ the 2D density of states enters the result. For spherically symmetric Fermi surface it does not depend on $k_z$.  Solution of Eq. (3) is searched in the form $\Sigma^{(2)}=(-i/2\tau)sgn \varepsilon$. Substitution in the Eq. (3) renders
$$
\frac{1}{\tau}=n_2 m^{\ast}\overline{|u|^2},                                      \eqno(4)
$$
where $\overline{|u|^2}=\int\frac{d\varphi'}{2\pi}|u(\kappa-\kappa')|^2$ is the transverse cross-section. Integration is going over directions of $\kappa'$ in the plane perpendicular to $z$. The absolute value of of $\kappa'$ depends on $k_z$ via  $|\kappa'|=\sqrt{k_F^2-k_z^2}$. The strands are assumed to be axially symmetric, then the cross-section does not depend on the incident $\kappa$. The resulting Green function is then
 $G(\omega_n,\mathbf{k})=[i(|\omega_n|+1/2\tau)sgn \omega_n -\xi(\mathbf{k})]^{-1}$

Important contributions of the same order originate from the terms describing processes of scattering of two excitations with the opposite momenta by the same impurity. The sum entering the r.h.s. of Eq. (2) can be rewritten as a series:
$$
\sum_n\int\frac{dk_z}{2\pi} k_z^2\int\frac{d^2\kappa}{(2\pi)^2}G(\omega_n,k_z,\kappa)G(-\omega_n,-k_z,-\kappa)[1+Q(\omega_n,k_z)+Q^2(\omega_n,k_z)+...], \eqno(5)
$$
where
$$
Q(\omega_n,k_z)=n_2\int\frac{d^2\kappa'}{(2\pi)^2}|u(\kappa-\kappa')|^2G(\omega_n,k_z,\kappa')G(-\omega_n,-k_z,-\kappa').          \eqno(6)
$$
A straightforward integration renders $Q(\omega_n,k_z)=\frac{1}{(2|\omega_n|\tau+1)}$ and for the sum of the geometric series in Eq. (5):
$S=(2|\omega_n|\tau+1)/2|\omega_n|\tau$. Since $\tau$ here does not depend on a direction of $\kappa$ the integration over $d^2\kappa$ in Eq. (5) results in the expression $\frac{m^{\ast}\tau}{(2|\omega_n|\tau+1)}\cdot\frac{(2|\omega_n|\tau+1)}{2|\omega_n|\tau}=\frac{m^{\ast}}{2|\omega_n|}$. This expression does not contain $\tau$ for all values of $k_z$ and $\omega_n$. Since  $\tau$ drops out of the equation for the
$T_c$ we can conclude, that mutually parallel and specularly reflecting strands do not lower the transition temperature of liquid $^3$He from the normal into the polar phase i. e. the phase, corresponding to the Cooper pairing with the orbital momentum $l=1$ and its projection on the direction of the strands $l_z=0$.
Formally  this result is a consequence of the same mutual compensation of different types of corrections to Eq. (1) as that, leading  to the Anderson theorem.
A physical analogy of the two problems is also obvious. Elastic scattering of excitations changes directions of their momenta but does not change their absolute values. For the $s$-wave pairing the transition temperature is determined by the spherically symmetric component of mutual interaction of excitations. It does not change in the scattering process. In case of the polar phase $T_c$ is determined by the  $3g(k_z\cdot k_z^{\prime})$ component of the interaction. At the specular reflection it also does not change.
\section{Other phases}
For the phases, corresponding to the pairing with $l_z=\pm 1$ the transverse  scattering has destructive effect and lowering of the $T_c$ is expected. The system of strands on the average is axially symmetric and $T_c$-s for both projections of $l_z$ are degenerate. It is sufficient to  consider  any linear combination of two. Let it be $k_y$. Eq. (3) does not contain an exolicit form of the order parameter, so that $\Sigma^{(2)}(\varepsilon,\kappa)$ does not change, but instead of the expression (6) for $Q(\omega_n,k_z)$  we have now an integral operator:
$$
\hat{Q}_{\bot}(\omega_n,k_z)k_y=n_2\int\frac{d^2\kappa'}{(2\pi)^2}|u(\kappa-\kappa')|^2k_y'G(\omega_n,k_z,\kappa')G(-\omega_n,-k_z,-\kappa').    \eqno(7)
$$
In agreement with the general procedure \cite{AG1} instead of the Born amplitude in this formula and in the definition of $1/\tau$ the full scattering amplitude  has to be substituted. Potential $u(\rho-\rho_a)$ is two-dimensional and the scattered wave is proportional to the Hankel function $H^{1}_0(kr)$. Diameters of the strands are much greater than the Fermi wave length, so that instead of the Hankel function its asymptotic expression for large values of the argument can be used. For the   $|u(k)|^2$ we substitute
$2\pi k_{\perp}\left(\frac{\hbar^2}{m^{*}}\right)^2|f(\varphi)|^2$, where $f(\varphi)$  is two-dimensional scattering amplitude i.e.  $|f(\varphi)|^2$ - two-dimensional cross-section, it has dimensionality of a length. The full cross-section is given by the integral  $\int_{-\pi}^{\pi}|f(\varphi)|^2d\varphi$. In the classical limit it is equal to diameter of a strand $2R$. With these simplifications the expression  (4) for  $1/\tau$ reads as
$$
\frac{1}{\tau}=n_2\frac{\hbar}{m^{\ast}}2R k_{\perp},                                       \eqno(8)
$$
as before $k_{\perp}$ depends on $k_z$: $ k_{\perp}=\sqrt{k_F^2-k_z^2}$.

Integration over directions of $\kappa'$ in the plane  normal to $z$ renders instead of the factor $1/\tau$ an expression  $-ck_y/\tau$ with a coefficient $c$, which depends on a character of scattering of excitations by a strand.  For the specular reflection by a circular cylinder $c$=1/3. The component $k_y$ is the eigenfunction of $\hat{Q}_{\bot}$:
$$
\hat{Q}_{\bot}k_y=-\frac{c}{(2|\omega_n|\tau+1)}k_y.                                                            \eqno(9)
$$
 The terms, containing powers of $(\hat{Q}_{\bot})$ form a geometric series.
Collecting all terms in a standard way we arrive at the equation for $T_{c\bot}$, which contains $\tau$:
$$
1=-3 g T_c\sum_{n\geq 0}\int\frac{d^3k}{(2\pi)^3} \hat{k_y}^2G(\omega_n\mathbf{k})G(-\omega_n\mathbf{-k})
\frac{2|\omega_n|\tau+1}{2|\omega_n|\tau+1+c}.        \eqno(10)
$$
Integration over the absolute value of $k$ in the r.h.s. of Eq. (10) renders:
$$
N(0)\pi\int\frac {do}{4\pi}\frac{\hat{k_y}^2}{|\omega_n|+(1+c)/(2\tau)},
$$
where $N(0)$ is the 3D density of states. In a contrast to isotropic impurities $1/\tau$ here depends on the direction of $\mathbf{k}$. If the polar axis is oriented in $z$-direction $1/\tau=\sin\theta/\tau_0$, where  $1/\tau_0=n_2\frac{\hbar}{m^{\ast}}2R k_F$. Because of the angular dependence of the denominator of the expression under the integral sign the integration here is not elementary. As a result the equation (10) does not have a usual form with the di-gamma function \cite{lark}. If $|\omega_n|\tau\gg 1$, i.e. effect of impurities is small, the fraction in the r.h.s. of Eq. (10) ca be expanded in powers of $(1+c)/(2|\omega_n|\tau)$ and integrated over $d\theta$ in every term. In the principal order over $1/|\omega_n|\tau$ we obtain:
$T_{c0}-T_{c1}=\frac{3\pi^2}{16}\frac{\hbar^2}{m^{\ast}}n_2Rk_F$,
where $T_{c0}$ and $T_{c1}$ - the temperatures of transition in the states with the projections of angular moment  $l_z=0$ and $l_z=\pm 1$ respectively.
 The global anisotropy induced by aerogel is phenomenologically described by additional term in the free energy, which is proportional to $\kappa_{jl}A_{\mu j}A_{\mu l}^*$, where $\kappa_{jl}$ is a real symmetric traceless tensor. If anisotropy is axial the principal values of this tensor are $\kappa,\kappa,-2\kappa$. Polar phase is realized at $\kappa<0$. The relative difference of critical temperatures $(T_{c0}-T_{c1})/T_{c0}=-3\kappa$  \cite{SF1,fom3}, but the real interval of stability of the polar phase is more wide because of the effect of the fourth order terms in the Landau expansion of the free energy. The transverse component of the order parameter appears at $T=T_{cA}$, which is different from $T_{c1}$:
 $$
 \frac{T_{c0}-T_{cA}}{T_{c0}}=\frac{3\pi^2}{16}\frac{\hbar^2}{m^{\ast}T_{c0}}n_2Rk_F\frac{\beta_{12345}}{2\beta_{13}},              \eqno(11)
 $$
where $\beta_1,... \beta_5$  are the coefficients in front of the fourth order terms in the Landau expansion of the free energy of superfluid $^3$He.  For  sums of these coefficients the short hand notations are used: $\beta_{12345}=\beta_1+\beta_3+...+\beta_5$, $\beta_{13}=\beta_1+\beta_3$. For the weak coupling values of these coefficients $\beta_{12345}/\beta_{13}=3$ and the interval of stability of the polar phase is
$\delta T/T_0=\frac{9\pi^2}{32}n_2\frac{\hbar^2}{m^{\ast}T_c}Rk_F$. This interval is of the order of $\xi_0/l_{\bot}$, where  $l_{\bot}$ is a characteristic mean free path in transverse direction. For finite $\delta T/T_0$ $T_{c1}$ can be found from  Eq.(10) expanded in a series over $1/T_c\tau$:
$$
\ln\left(\frac{T_{c0}}{T_{c1}}\right)=\frac{3}{2}\sum_{m=1}^{\infty}J_m\left(\frac{1}{3\pi T_{c1}\tau_0}\right)^m\frac{1}{m!}\psi^{(m)}(1/2).  \eqno(12)
$$
Here $\psi^{(m)}(1/2)$ is the $m$-th derivative of the di-gamma function, and $J_m=\int_0^{\pi}(\sin\theta)^m d\theta=\sqrt{\pi}\frac{\Gamma(2+m/2)}{\Gamma(2+(m+1)/2)}$.

\section{Magnetic scattering}
If strands of nafen are not covered by the sufficiently thick film of $^4$He a layer of solid $^3$He is formed on their surface. This influences scattering of excitations by the strands. The transition from the specular reflection to the diffuse decreases the global anisotropy. A quantitative analysis of this decrease requires substantial modification of the model, it will be not carried out here. We consider here only the mechanism of suppression of $T_{c}$ by a magnetic scattering.\, assuming that it is dominant. For an estimation of the contribution of magnetic scattering we apply the argument of the AG theory \cite{AG3}. The paramagnetic solid $^3$He can exchange spins with quasi-particles and break the Cooper pairs like magnetic impurities in conventional superconductors.  Interaction of spins of quasi-particles with the spins of $^3$He atoms is described by the exchange Hamiltonian:
$$
H_{int}=\sum_a J\psi^{\dag}(\mathbf{r_a})\hat{\sigma}^k\hat{S}^k_a\psi^{\dag}(\mathbf{r_a}).                        \eqno(13)
$$
Here $\hat{S}^k_a$ is the operator of the spin of the impurity with the coordinate $\mathbf{r_a}$. For an estimation it is sufficient to assume that spins are  randomly distributed with the average density $n_s$ and randomly oriented, then we can literally follow the argument of AG-theory \cite{AG3}. Difference with respect to the s-wave case consists in the spin structure of the order parameter.  For the singlet pairing  $\Delta_{\alpha \beta}\sim (i\sigma_y)_{\alpha \beta}$, but for the triplet Cooper pairing, which is the case for $^3$He it is  $\Delta_{\alpha \beta}\sim\vec{d}(i\vec{\sigma}\sigma_y)_{\alpha \beta}$. This difference does not cause a serious difficulty if one assumes that $\vec{d}=(1,0,0)$ i.e. has only one component. Within the assumption of random distribution of impurities this does not restrict generality of the argument. The resulting equation for $T_c$ does not change its form:
$$
\ln\frac{T_{c0}}{ T_{cs}}=\psi\left(\frac{1}{2}+\frac{1}{2\pi\tau_s T_{cs}}\right)-\psi\left(\frac{1}{2}\right),         \eqno(14)
$$
where
$$
\frac{1}{\tau_s}=\frac{\pi N_0n_sJ^2}{4}.                                                              \eqno(15)
$$
For a small concentration of impurities the decrease of the $T_c$ is proportional to the concentration:
$$
T_{c0}-T_{cs}=\frac{\pi}{4\tau_s}.                               \eqno(16)
$$
This decrease can be compared with the decrease (11) caused by the transverse scattering of excitations. For the estimation of $n_s$ we assume that the adsorbed atoms of $^3$He form a monolayer on the surface of nafen. It takes about 10 square  Angstroms of surface per one atom. Then from Eqns. (11),(15),(16)
$$
\frac{T_{c0}-T_{cs}}{T_{c0}-T_{c1}}\sim (N_0 J)^2.                            \eqno(17)
$$
Using the estimation $J\approx$ 100 mK, from the Ref.\cite{coll} we arrive at  $(N_0 J)^2\approx 1/10 \div 1/20$. According to this estimation magnetic scattering is not dominating mechanism  of the observed suppression of the polar phase and relevant changes of the phase diagram of  $^3$He.

\section{Conclusion}
The above analysis renders a qualitative explanation of the observed sensitivity of the phase diagram of liquid $^3$He in nafen to the coverage of its strands by a film of $^4$He.  Because of the film the scattering of the Fermi excitations by the strands is close to the specular. For such scattering the longitudinal with respect to the strands component of momentum is conserved. In this case, at a strength of the result of Sec.2 the temperature of transition from the normal into the polar phase remains the same as that of the transition from the normal into any of the superfluid phases with $l$=1 in the bulk  $^3$He. Temperatures of transitions in possible phases with $l_z=\pm 1$ are depressed for a finite value because of the scattering of excitations  by the strands of nafen. As a result in a finite interval of temperatures only a phase with $l_z=0$, i.e. the polar phase is stable. At further lowering of temperature distortions of the order parameter occur due to the admixture of components with projections $l_z=\pm 1$.

If a film of $^4$He is absent the scattering of excitations by the strands is diffuse. That removes a qualitative difference between different projections of moments of excitations. Another mechanism of rapprochement  of the transition temperatures into the phases with different projections of $l$ is the exchange of spins of Fermi excitations with the spins of the adsorbed on the strands atoms of  $^3$He. Both factors bring the system closer to the state when the choice of phases is determined not by the arrangement of impurities, but by the terms of the 4-th order in the Landau expansion of the free energy.

The results of Sec.2 and Sec.3 do not  apply  to aerogels with randomly oriented strands. That explains why the phase diagram of  $^3$He in these aerogels is not so sensitive to coverage of the strands by a film of  $^4$He (cf. Ref.\cite{Dm-ms}). Scattering by a randomly oriented strands does not select one of the projections of momenta both for the specular and for the diffuse boundary conditions. The effect of magnetic scattering, as follows from the estimation, made in Sec.4, is not dominating.

\section{Acknowledgements}
This paper is written for the special issue of JETP, devoted to the 85-th anniversary of Lev Petrovich Pitaevskiy. I had luck to be his PhD student and during this period and for many years afterwards to have possibility to discuss with him scientific and not only scientific questions. I am happy to use the opportunity to thank him for that and to wish him a good health and continuation of his outstanding scientific activity. 

I thank V.V. Dmitriev for useful discussions and constructive criticism. The work is supported in part by the Program of Presidium of RAS 1.1 "Current problems of the low temperature physics".

\end{document}